\documentclass[11pt]{article}
\usepackage{latexsym}  
\usepackage{amssymb}
\usepackage{graphicx}
\usepackage{amsmath}

\begin{document}

\begin{center}

{\Large\bf Quark and Lepton Mass Matrices} \\[2mm]

{\Large\bf Described by Charged Lepton Masses}

\vspace{4mm}

{\bf Yoshio Koide$^a$ and Hiroyuki Nishiura$^b$}

${}^a$ {\it Department of Physics, Osaka University, 
Toyonaka, Osaka 560-0043, Japan} \\
{\it E-mail address: koide@kuno-g.phys.sci.osaka-u.ac.jp}

${}^b$ {\it Faculty of Information Science and Technology, 
Osaka Institute of Technology, 
Hirakata, Osaka 573-0196, Japan}\\
{\it E-mail address: hiroyuki.nishiura@oit.ac.jp}

\date{\today}
\end{center}

\vspace{3mm}

\begin{abstract}
Recently, we proposed a unified mass matrix model for quarks and leptons, 
in which, mass ratios and mixings of 
the quarks and neutrinos are described by using only the 
observed charged lepton mass values as family-number-dependent 
parameters  and only six family-number-independent free 
parameters.  In spite of  quite few parameters,
the model  gives remarkable agreement with observed data (i.e. 
CKM mixing, PMNS mixing and mass ratios).
Taking this phenomenological success seriously, we give a formulation
of the so-called Yukawaon model  in details from a theoretical aspect, 
especially for the construction of superpotentials and $R$ charge 
assignments of fields. 
The model is considerably modified from the previous one, while 
the phenomenological success is kept unchanged.   
\end{abstract}

PCAC numbers:  
  11.30.Hv, 
  12.15.Ff, 
  14.60.Pq,  
  12.60.-i, 

\vspace{3mm}

{\large\bf 1  \  Introduction} 
 
It is a big concern in the flavor physics to investigate the origin of 
the observed hierarchical structures of  masses 
and mixings of quarks and leptons.
Recently, a unified mass matrix model for quarks and leptons 
was proposed \cite{Yukawaon_seesaw}:  In the model,  
mass ratios and mixings of the quarks and neutrinos are 
described by using only the observed charged lepton mass 
values as ``family-number-dependent" parameters  
and only six ``family-number-independent" free parameters.  
In spite of  quite few parameters, the model  gives remarkable 
coincidence with observed all data, i.e. Cabibbo-Kobayashi-Maskawa 
(CKM) mixing \cite{CKM_1,CKM_2} in quark sector and 
 Pontecorvo-Maki-Nakagawa-Sakata (PMNS) \cite{PMNS_1,PMNS_2}
mixing in lepton sector,  and quark and lepton mass ratios. 
Besides, the model gives very interesting predictions 
for leptonic $CP$ violation parameter 
$\delta_{CP}^\ell \simeq -70^\circ \simeq - \delta_{CP}^q$ 
and effective Majorana neutrino mass $\langle m\rangle\simeq 21$ meV. 
We list those numerical results in Table 1, which was quoted from
Ref.\cite{Yukawaon_seesaw}.

\begin{table}
\caption{Predicted values vs. observed values. [Quoted from  
\cite{Yukawaon_seesaw}]. 
The predicted values are obtained only from inputs of the six 
family-number-independent parameters $b_u=-1.011$, 
$b_d= -3.3522$, $\beta_d =17.7^\circ$, 
$(\tilde{\phi}_1, \tilde{\phi}_2 ) =(-176.05^\circ, -167.91^\circ)$, 
and $\xi_R = 2039.6$.  The observed values were quoted from 
Ref.\cite{PDG14}.}

\vspace*{2mm}
\hspace*{-17mm}
\begin{tabular}{|c|ccccccccc|} \hline
  & $|V_{us}|$ & $|V_{cb}|$ & $|V_{ub}|$ & $|V_{td}|$ & 
$\delta^q_{CP}(^\circ)$ &  $r^u_{12}$ & $r^u_{23}$ & $r^d_{12}$ & $r^d_{23}$ 
 \\ \hline 
Predicted &$0.2257$ & $0.03996$ & $0.00370$ & $0.00917$ & $81.0$ & 
$0.061$ & $0.060$ & $0.049$ & $0.027$ 
 \\
Observed & $0.22536$ & $0.0414$ &  $0.00355$  & $0.00886$  & $69.4$ &
$0.045$ & $0.060$ & $0.053$  & $0.019$  
  \\ 
    &  $ \pm 0.00061$ &  $ \pm 0.0012$& $ \pm 0.00015$ & 
 ${}^{+0.00033}_{-0.00032}$ & $\pm 3.4$ &
${}^{+0.013}_{-0.010}$ & $ \pm 0.005$ & $^{+0.005}_{-0.003}$ &
${}^{+0.006}_{-0.006}$ 
 \\ \hline
   & $\sin^2 2\theta_{12}$ & $\sin^2 2\theta_{23}$ & $\sin^2 2\theta_{13}$ & 
 $R_{\nu}\ (10^{-2})$ &  
$\delta^\ell_{CP}(^\circ)$ & $m_{\nu 1}\ ({\rm eV})$ & $m_{\nu 2}\ ({\rm eV})$ & 
$m_{\nu 3}\ ({\rm eV})$ & $\langle m \rangle \ ({\rm eV})$ \\ \hline
 Predicted & $0.8254$ & $0.9967$ & $0.1007$ &  $3.118$ & $-68.1$ &
 $0.038$ & $0.039$ & $0.063$ & $0.021$ \\
Observed & $0.846$   & $ 0.999$ & $0.093$ &   $3.09 $    &no data
  &no data  &no data  &no data  &  $<\mathrm{O}(10^{-1})$   \\ 
    & $ \pm 0.021$ & $^{+0.001}_{-0.018}$   & $\pm0.008$ & $ \pm 0.15 $  &  &
   &    &    &    \\ \hline 
\end{tabular}
\end{table}

The previous paper has focused on only phenomenological 
aspect of the model and shown the phenomenological success
which should be taken seriously. 
However, discussions on the theoretical aspect of the model 
was somewhat not sufficient. 
So, in this paper, we shall give a formulation of a new Yukawaon 
model from the theoretical aspect. 

In the so-called Yukawaon model \cite{Yukawaons_1,Yukawaons_2,Yukawaons_3,
Yukawaons_4,Yukawaons_5,Yukawaons_6,Yukawaons_7,Yukawaons_8,
Yukawaons_9,Yukawaons_10}, we regard 
Yukawa coupling constants $(Y_f)_{ij}$ ($f=u,d,\nu, e)$ as 
effective quantities $(Y_f^{eff})_{ij}$ which are given by 
vacuum expectation values (VEVs) of scalares $Y_f$ as 
$(Y_f^{eff})_{ij} = y_f \langle (Y_f) \rangle_{ij}/\Lambda$.
The model is a sort of flavon model \cite{flavon,flavon_2}. 
(For recent works, for instance, see \cite{flavon_3,flavon_4}.)
The model is based on family symmetries U(3)$\times$U(3)$'$. 
The U(3)$\times$U(3)$'$ symmetries are broken at $\mu=\Lambda$ 
and $\mu=\Lambda'$. 
(We assume $\Lambda \ll \Lambda'$.) 
The symmetry U(3)$'$ is broken into $S_3$ at an energy scale $\Lambda'$ 
through the vacuum expectation values (VEVs) of flavons 
$(S_f)_\alpha^{\ \beta}$ which  take a form ``unit matrix ${\bf 1}$ 
plus democratic matrix $X_3$", 
$\langle S_f \rangle = v_{Sf}( {\bf 1} + b_f X_3)$. 
Here, $f$ is sector names $f= u, d, \nu, e$, 
and  indices $i$, $j$ and $\alpha$, $\beta$ are those in
U(3) and U(3)$'$, respectively. 
(For the details, see Eq.(2.7) later. )
The flavons $S_f$ play an essential role as we discuss in Eq.(2.3). 
The parameters $b_f$ are  typical ``family-number 
independent parameters" in the Yukawaon model, and they 
 determine not only mass spectrum of the fermion 
$f$ but also mixing among $f_i$. 
On the other hand, the U(3) family symmetry is completely broken
by VEV of a flavon $\Phi_0$ at $\mu= \Lambda$ as we discuss in Eq.(2.10) later. 
We do not consider any subgroups of U(3).
Instead, from the practical point of view, we use the 
observed charged lepton masses for inputs of the VEV 
$\langle \Phi_0 \rangle \equiv v_0 {\rm diag} (z_1, z_2, z_3)$,
which are only ``family-number dependent parameters" in the 
Yukawaon model.  
We do not ask the origin of the values of $z_i$ in this
paper.
This is future task in our investigation.
Furthermore, the last basic hypothesis in the Yukawaon model 
is that the VEV forms of flavons are diagonal except for
$\langle S_f \rangle$ which take the S$_3$ invariant forms. 
The details are discussed in the next section.

The VEV relations in the model are derived from superpotentials 
which are invariant 
under the  U(3)$\times$U(3)$'$ and constructed by using suitable 
$R$ charge assignments. 
Once we give superpotential form invariant under U(3)$\times$U(3)$'$
with $R$ charge conservation, we can uniquely obtain our desirable
VEV relation as seen in Sec.3. 
Therefore, the $R$ charge assignment is crucial for the phenomenological success.  
However, 
the explicit forms of superpotentials and $R$ charge 
assignments were not presented in the previous paper 
\cite{Yukawaon_seesaw}. 
The purpose of the present paper is to give explicit 
superpotential forms and $R$ charge assignments in details. 
In this paper, 
we obtain more natural $R$ charge assignment than that 
in the previous model \cite{Yukawaon_seesaw}. 
The new $R$ charges assignment is given in Table 2. 
This assignment causes a change of the formulation given in the previous 
paper, including a new relation among the phase parameters in VEV of a flavon $P$ and the observed charged lepton masses. 

A flavon $P_u$ presented later in (2.13) with VEV of phase matrix type 
$\langle P \rangle = {\rm diag}(e^{i\phi_1}, e^{i\phi_2}, e^{i\phi_3})$ 
plays an essential role in the Yukawaon model and
the phases $(\phi_1, \phi_2, \phi_3)$ are described in term of 
charged lepton mass values \cite{K-N_PRD15}. 
In the the previous model, we did not discuss the explicit 
mechanism but referred to a mechanism given in Ref.~\cite{K-N_PRD15}, which cannot be
straightforwardly applied to the model \cite{Yukawaon_seesaw}
because the $R$ charge assignment in the model \cite{Yukawaon_seesaw}
is different from that in the model \cite{K-N_PRD15}.
The new $R$ charge assignment in the Table 2 is also different from 
\cite{K-N_PRD15} and  \cite{Yukawaon_seesaw}, so that 
we are led to a new relation among the phase parameters and 
the observed charged lepton masses. 
As a result, we obtain different parameter values $(\phi_1, \phi_2,\phi_3)
\equiv (\tilde{\phi}_1 + \phi_0, \tilde{\phi}_2 + \phi_0, \phi_0)$, 
while,  as far as the values $(\tilde{\phi}_1, \tilde{\phi}_2)$ are
concerned, we can obtain the same values as those in the previous paper 
\cite{Yukawaon_seesaw}.  
(However, this does not mean that present model is identical with 
the previous one \cite{Yukawaon_seesaw}.  Note that,  
in the CKM fitting, only the values $(\tilde{\phi}_1, \tilde{\phi}_2)$ 
are observable and $\phi_0$ is not observable, while the parameter $\phi_0$ 
is observable in the U(3) family model. )  
The details are given in Sec.4.

The present paper is arranged as follows:
In Sec.2, the basic postulation in the Yukawaon model 
\cite{Yukawaons_1,Yukawaons_2,Yukawaons_3,
Yukawaons_4,Yukawaons_5,Yukawaons_6,Yukawaons_7,Yukawaons_8,
Yukawaons_9,Yukawaons_10}and the VEV relations in the previous paper 
\cite{Yukawaon_seesaw} are reviewed without 
showing the explicit superpotentials. 
In Sec.3, we will discuss superpotentials which give  
special VEV forms playing an essential role in the 
phenomenological investigation. 
In Sec.4, we will discuss the relation of phase parameters 
defined by Eq.(2.13) to the charged lepton masses. 
Finally, Sec.5 is devoted to summary and concluding remarks.

\vspace{3mm}

{\large\bf 2 \ Basic assumptions in the Yukawaon model and its VEV relations} 

We  investigate flavor physics from the point of
view of family symmetry.
It is unnatural that the Yukawa coupling constants $Y_f$ explicitly break 
the family symmetry. 
Therefore, in order that the Hamiltonian is invariant under the symmetry,
we must consider that $Y_f$ are effective coupling constants 
$Y_f^{eff}$ 
which are given by 
vacuum expectation values (VEVs) of scalars (``Yukawaons" 
\cite{Yukawaons_1,Yukawaons_2,Yukawaons_3,
Yukawaons_4,Yukawaons_5,Yukawaons_6,Yukawaons_7,Yukawaons_8,
Yukawaons_9,Yukawaons_10})  
$Y_f$ with $3 \times 3$ components for each sector $f$:
$$
(Y_f^{eff})_i^{\ j} =  \frac{y_f}{\Lambda}  \langle Y_f\rangle_i^{\ j}  
\ \ \ \ (f=u, d, \nu, e),
\eqno(2.1)
$$
where $\Lambda$ is an energy scale of the effective theory. 
All the flavons in the Yukawaon model are expressed 
by $3 \times 3$ components. 
Would-be Yukawa interactions are given by 
$$
H_Y = \frac{y_\nu}{\Lambda} (\bar{\ell}_L)^i (\hat{Y}_\nu)_i^{\ j}
 (\nu_R)_j  H_u 
+ \frac{y_e}{\Lambda} (\bar{\ell}_L)^i (\hat{Y}_e)_i^{\ j} (e_R)_j  H_d 
+ y_R (\bar{\nu}_R)^i (Y_R)_{ij} (\nu_R^c)^j 
$$
$$
+ \frac{y_u}{\Lambda}  (\bar{q}_L)^i (\hat{Y}_u)_i^{\ j} (u_R)_j H_u 
+ \frac{y_d}{\Lambda}  (\bar{q}_L)^i (\hat{Y}_d)_i^{\ j}  (d_R)_j H_d  ,
\eqno(2.2)
$$
where we have assumed a U(3) family symmetry, and 
$\ell_L=(\nu_L, e_L)$ and $q_L=(u_L, d_L)$ are SU(2)$_L$ doublets. 
$H_u$ and $H_d$ are two Higgs doublets. 
Those Yukawaons $\hat{Y}_f$ are distinguished from each other  
by $R$ charges.
Hereafter, for convenience, we use notations 
$\hat{A}$, $A$, and $\bar{A}$ for fields with ${\bf 8}+{\bf 1}$,
${\bf 6}$, and ${\bf 6}^*$ of U(3), respectively.  
In addition, another types of flavons $A_i^{\ \alpha}$ and 
$A_{i\alpha}$ appear in the present U(3)$\times$U(3)$'$ model.
Since we pay attention only to the index of U(3), we denote
anti-flavons of those as $\bar{A}^i_{\ \alpha}$ and 
$\bar{A}^{i\alpha}$, respectively.    

In the present model, we have flavons $(\hat{Y}_f)_i^{\ j}$, 
$(\Phi_{0f})_i^{\ \alpha}$, $(P_f)_i ^{ \alpha}$, 
$({S}_f)_\alpha^{\ \beta}$, and $(\Theta_{0f})_i^{\ \alpha}$. 
We assume that VEV matrices of those flavons take diagonal 
forms (except for $({S}_f)_\alpha^{\ \beta}$ which take S$_3$ 
invariant forms) at our basic flavor basis as we discuss later. 
In addition to those flavons, we consider other flavons $(\Phi_0)_{ij}$,
$(\bar{Y}_R)^{ij}$, $(\Theta_R)_{ij}$, $E_{ij}$, $\hat{E}_i^{\ j}$,
and $\hat{\Theta}_\phi$.
The transformation properties and the $R$ charges of 
those flavons are listed in Table 2.
A role of each flavon will be discussed step by step below.

\vspace{2mm}
\begin{table}
\caption{Transformation properties and $R$ charges of flavons, 
quarks, leptons, and Higgs scalars in the present model. 
In the table, we omit row on U(3)$'$ for quarks, leptons 
and Higgs scalars,  since it is 
obvious that those are singlets of U(3)$'$.
Also, we omit row on SU(2) for other flavons, because it is obvious 
that those are SU(2) singlets.  
We always consider a flavon 
$\bar{A}$ correspondingly to a flavon $A$.  
}
\vspace{2mm}
\begin{center}
\begin{tabular}{|c|ccc|ccc|cc|} \hline
quark, lepton, Higgs & $\ell$ & $\nu^c$ &  $e^c$ & 
$q$ & $u^c$  &   $d^c$  &  $H_u$  &  $H_d$  \\ \hline
SU(2)                &  {\bf 2}  &   {\bf 1} & {\bf 1} &
 {\bf 2}  &   {\bf 1} & {\bf 1} &  {\bf 2} &  {\bf 2} \\ 
U(3)                 &  {\bf 3}  & $ {\bf 3}^*$ &   $ {\bf 3}^*$ &
{\bf 3}  & $ {\bf 3}^*$ &   $ {\bf 3}^*$ &  {\bf 1} & {\bf 1} \\
$R$ charge & $\frac{5}{4}$ & $-\frac{1}{4}$ & $-\frac{1}{4}$ & 
0  & $\frac{1}{2}$ & $+\frac{3}{2}$ & 0 & 0 \\ \hline
\end{tabular}
\begin{tabular}{|c|cccc|cccc|cccc|} \hline
flavon & $\hat{Y}_e$ &  $\hat{Y}_\nu$ & $\hat{Y}_d$ & $\hat{Y}_u$  &   
$\Phi_{0e}$  &  $\Phi_{0\nu}$  &  $\Phi_{0d}$  &   $\Phi_{0u}$  &
$P_e$ &  $P_\nu$ &  $P_d$ &$P_u$  \\ \hline
U(3)      & ${\bf 8+1}$  & ${\bf 8+1}$ &  ${\bf 8+1}$  &   ${\bf 8+1}$  & 
${\bf 3} $  & ${\bf 3} $  &  ${\bf 3} $ &  ${\bf 3} $  &   ${\bf 3} $   & 
${\bf 3} $ & ${\bf 3} $ &  ${\bf 3} $   \\
U(3)$'$   &   ${\bf 1} $  &   ${\bf 1} $    & ${\bf 1} $  & ${\bf 1} $  &  
 ${\bf 3}^*$ & ${\bf 3}^* $   &  ${\bf 3}^* $  &  ${\bf 3}^* $  & 
 ${\bf 3} $  & ${\bf 3} $  &  ${\bf 3} $  &  ${\bf 3} $  \\
$R$ charge &   1  &  1   &  $\frac{1}{2}$   &  $\frac{3}{2}$ & 
1  &  1 &  $\frac{1}{2}$   &  $\frac{3}{2}$ &
 $\frac{1}{2}$  &  $\frac{1}{2}$ & $0$ & $1$  
\\ \hline
 \end{tabular}
 \begin{tabular}{|cccc|cccc|ccccc|} \hline
   $\hat{S}_e$   &  $\hat{S}_\nu$  &  $\hat{S}_d$ & $\hat{S}_u$
  & $\Theta_{0e}$  &  $\Theta_{0\nu}$  &  $\Theta_{0d}$  &   $\Theta_{0u}$ &
$\Phi_0$ &  $E$ &  $\bar{Y}_R$ &$\Theta_R$ & $\hat{\Theta}_\phi$ 
\\ \hline
 ${\bf 1}$ &  ${\bf 1}$  &   ${\bf 1}$  &  ${\bf 1}$ &
${\bf 3} $  & ${\bf 3} $  &  ${\bf 3} $ &  ${\bf 3} $  &   
${\bf 6} $   & ${\bf 6} $ &  ${\bf 6} ^*$ &  
${\bf 6} $ & ${\bf 8+1}$  \\
 ${\bf 8+1}$ & ${\bf 8+1}$  &  ${\bf 8+1}$  &  ${\bf 8+1}$  & 
 ${\bf 3}^* $  & ${\bf 3}^* $  &  ${\bf 3}^* $  &  ${\bf 3}^* $ & 
 ${\bf 1}$ &  ${\bf 1}$ &  ${\bf 1}$ & ${\bf 1}$ &  ${\bf 1}$  \\
 1  &  1   &   $\frac{1}{2}$   &  $\frac{3}{2}$ &  
1  &  1   &   $\frac{3}{2}$  &   $\frac{1}{2}$  & 
 $\frac{1}{2}$ &  $\frac{1}{2} $  &  $\frac{5}{2}$  
&  $-\frac{1}{2}$  & 1 \\
 \hline
\end{tabular}
\end{center}
\end{table}

\vspace{2mm}

Let us list VEV relations which are our goal: 
The following (i)-(v) are used in the previous model 
\cite{Yukawaon_seesaw}, while (vi) is revised from the previous paper.
We will derive these relations from superpotentials presented in Sec.3
in this paper.

(i) The VEV of the Yukawaons $\langle \hat{Y}_f \rangle$ are 
given by the following relations:
$$
\langle \hat{Y}_f \rangle_i^{\ j} = k_f \ 
 \langle {\Phi}_{0f} \rangle_i^{\ \alpha} 
\langle (S_f)^{-1} \rangle_{\alpha}^{\ \beta} 
\langle \bar{\Phi}_{0f}^T \rangle_{\beta}^{\ j} 
\ \ \ \ (f=u, d, \nu, e) .
\eqno(2.3)
$$
The factor $S_f^{-1}$ in Eq.(2.3) comes from a seesaw-like scenario 
by assuming new heavy fermions $F_\alpha$ and considering 
the following $6\times 6$ mass matrix model:
$$
(\bar{f}_L^i \ \ \bar{F}_L^\alpha ) 
\left(
\begin{array}{cc}
(\hat{Y}_f)_i^{\ j}  &  (\Phi_{0f})_i^{\ \beta}  \\
(\bar{\Phi}_{0f}^T)_\alpha^{\ j} & -(S_f)_\alpha^{\ \beta} 
\end{array} \right) 
 \left(
\begin{array}{c}
f_{Rj} \\
F_{R\beta}
\end{array} 
\right) .
\eqno(2.4)
$$
Here $f_{L(R)}$ and $F_{L(R)}$ are, respectively, 
left (right) handed light and heavy  fermions fields. 
Exactly speaking, we have to read $\bar{f}_L$ in Eq.(2.4) as 
$\bar{f}_L H_{u/d}/\Lambda$. 
However, for convenience, we have denoted
those as $\bar{f}_L$ simply.
Such a seesaw-like scenario with the democratic form of $S_f$ 
has been proposed by Fusaoka and one of the authors (YK) 
\cite{YK-HF_ZPC96} in order to understand the observed fact
$$
m_t \sim {\Lambda_{weak}}, \ \ \ \ \ \ 
m_u \sim m_d \sim m_e .
\eqno(2.5)
$$
In the Yukawaon model, 
when we consider $|\hat{Y}_f| \ll |\Phi_{0f}| \ll |S_f|$ 
(i.e. $\Lambda \ll \Lambda'$), we obtain a mass matrix for
$\bar{f}'_L$ and $f'_R$, 
$$
M_f \simeq \hat{Y}_f + \Phi_{0f} S_f^{-1} \bar{\Phi}_{0f} ,
\eqno(2.6)
$$
after the block diagonalization of Eq.(2.4). 
Regrettably, this relation (2.6) is not one we want, 
because the first term $\hat{Y}_f$ in Eq.(2.6) is independent 
of the second term $ \Phi_{0f} S_f^{-1} \bar{\Phi}_{0f}$. 
Therefore, in the previous paper \cite{Yukawaon_seesaw}, 
we have put some phenomenological assumptions in order to 
obtain the relation (2.3).  
We are still not satisfied with the scenario given in the 
previous paper \cite{Yukawaon_seesaw}, and we think that 
the scenario should be improved. 
However, for simplicity, in the present paper, we still 
denote Dirac mass matrices $M_f$ as Eq.(2.3).

(ii) The VEV form of $\langle {S}_f \rangle$, which is due to 
the symmetry breaking U(3)$' \rightarrow\, $S$_3$,  is given by 
$$
\langle \hat{S}_f \rangle = v_{Sf} ({\bf 1} + b_f X_3) ,
\eqno(2.7)
$$
where ${\bf 1}$ and $X_3$ are defined as
$$
{\bf 1} = \left( 
\begin{array}{ccc}
1 & 0 & 0 \\
0 & 1 & 0 \\
0 & 0 & 1 
\end{array} \right) , \ \ \ \ \ 
X_3 = \frac{1}{3} \left( 
\begin{array}{ccc}
1 & 1 & 1 \\
1 & 1 & 1 \\
1 & 1 & 1 
\end{array} \right) . 
\eqno(2.8)
$$
The parameters $b_f$ in Eq.~(2.7) are typical examples of 
``family-number-independent" parameters. 
However, we will be obliged to take $b_e =0$ and $b_\nu=0$ 
in the lepton sector $f=e$ and $f=\nu$ as seen in Eq.(3.1) later. 
Therefore, 
the Dirac mass matrices $M_e$ and $M_\nu$ take diagonal forms 
without $b_f$ parameters (except for parameters in 
$\langle \Phi_0\rangle$ (see Eq.(2.10)), and quark mass matrices
$M_u$ and $M_d$ are described only by parameters 
$b_u$ and $b_d$, respectively. 

Although we have used the discrete symmetry S$_3$, our aim 
in the Yukawaon model is to understand all mass spectra 
and mixing on the basis of U(3)$\times$U(3)$'$ symmetries
without introducing any additional subgroups, e.g. S$_2$,  
A$_4$, SU(2), and so on.
If we adopt such symmetries, we will be obliged to accept
unwelcome family-number dependent parameters. 
This is against with the aim of Yukawaon model. 
Instead, from the practical point of view, we use only 
the observed charged lepton masses 
$m_{ei}$ as a result of U(3) symmetry breaking as seen 
in Eq.(2.10).

(iii) Motivated by the relation given below in (2.11), we assume that 
the VEV forms $\langle \Phi_{0f}  \rangle$ are diagonal in the 
flavor basis in which $\langle S_f \rangle$ take the forms (2.7),
and are given by
$$
\langle \Phi_{0f}  \rangle_i^{\ \alpha} = k_{0f} 
\langle \Phi_0 \rangle_{ik} 
\langle \bar{P}_f \rangle^{k\alpha}.
\eqno(2.9)
$$
 
The VEV of a flavon $\Phi_0$, $\langle \Phi_{0}  \rangle$ 
was defined by
 $$
\langle \Phi_0 \rangle = v_0 \, {\rm diag}\, (z_1, z_2, z_3)
 \propto {\rm diag}\, 
(\sqrt{m_e}, \sqrt{m_\mu},  \sqrt{m_\tau}), 
\eqno(2.10)
$$
where $z_i$ is normalized as $z_1^2+z_2^2+z_3^2=1$. 
The $\langle \Phi_{0}  \rangle$ plays a crucial role in the 
phenomenological investigation of the Yukawaon model.  
The existence of such the VEV matrix $\langle\Phi_0\rangle$ was 
suggested by a phenomenological success of the charged lepton mass
relation \cite{K-mass_1,K-mass_2,K-mass_3}.  
(For a recent work, for see \cite{K-mass_4}.)
$$
K \equiv \frac{m_e + m_\mu + m_\tau }{
(\sqrt{m_e} + \sqrt{m_\mu} + \sqrt{m_\tau} )^2}
= \frac{2}{3},
\eqno(2.11)
$$
which is excellently satisfied by the observed charged 
lepton masses (pole masses) as 
$K=(2/3)\times (0.999989\pm 0.000014)$. 
If we accept the flavon $\Phi_0$ with its VEV  (2.10), 
the charged lepton relation (2.11) is simply expressed as
$$
K = \frac{{\rm Tr}[\langle\Phi_0\rangle \langle\Phi_0]}{ 
({\rm Tr} [\langle\Phi_0\rangle])^2}.
\eqno(2.12)
$$
However, the purpose of the Yukawaon model is not to understand 
the relation (2.11). Therefore, 
in the Yukawaon model, we do not ask the origin of the charged 
lepton mass spectrum. 
It is a future task to understand the origin of the mass values 
$(m_e, m_\mu, m_\tau)$. 
In this paper we accept the observed  charged lepton mass values 
as fundamental family-dependent parameters, and we give 
a unified description of
quark and lepton mass matrices. 

On the other hand, we know that there exist the CKM mixing 
in quark sector and the PMNS mixing in lepton sector. 
Therefore, true regularity in the mass spectra ought to be disturbed  
by such mixings. 
The relation (2.11) is a specific case only for the charged leptons. 
We consider that a fundamental flavor basis 
in flavor physics is a basis in which 
the charged lepton mass matrix is diagonal.  
Moreover, we speculate that all masses and mixings of 
quarks and leptons might be described by inputting 
the observed charged lepton mass values. 
Under this idea, the Yukawaon model has 
been introduced and investigated \cite{Yukawaons_1,Yukawaons_2,
Yukawaons_3,Yukawaons_4,Yukawaons_5,Yukawaons_6,Yukawaons_7,
Yukawaons_8,Yukawaons_9,Yukawaons_10}.

(iv) The VEV form  $\langle P_f \rangle$ are defined as 
$$
\langle P_u  \rangle= v_P \, {\rm diag} (e^{i \phi_1},\ e^{i \phi_2},\, 
e^{i \phi_3}),
\ \ \ \langle  P_d \rangle = v_P {\bf 1} ,  \ \ \ 
\langle  P_\nu  \rangle= v_P {\bf 1}, \ \ \ 
\ \ \ \langle  P_e \rangle = v_P {\bf 1} .
\eqno(2.13)
$$
As we show $\langle P_f \rangle \langle \bar{P}_f \rangle = {\bf 1}$ 
in Sec.3, the special choices (2.13) are obviously ansatzes. 
(Also see a comment below Eq.(3.10) later.)
The parameters $(\phi_1, \phi_2, \phi_3)$  in Eq.~(2.13)
are typical examples of ``family-number-dependent" 
parameters.
However, in Sec.4, we will show that the parameters 
$(\phi_1, \phi_2, \phi_3)$ are described by the 
charged lepton masses $(m_e, m_\mu. m_\tau)$ 
with help of two family-number-independent parameters.
The origin of these VEV forms will be discussed in Sec.3.

(v) A neutrino mass matrix is given as
$$
(M_\nu^{Majorana})_{ij} =\langle \hat{Y}_\nu \rangle_i^{\ k} 
\langle \bar{Y}_R^{-1} \rangle_{kl} 
\langle \hat{Y}_\nu^T \rangle^l_{\ j} ,
\eqno(2.14) 
$$
by adopting the conventional 
seesaw mechanism \cite{seesaw_1,seesaw_2,seesaw_3,seesaw_4}.  
Here in this paper, 
differently from the previous paper \cite{Yukawaon_seesaw}, 
we assume the following VEV structure of the $Y_R$ (the Majorana mass 
matrix of the right-handed neutrinos $\nu_R$):  
$$
 \langle \bar{Y}_R \rangle^{ij} =k_R  \left[  
\left( \langle \bar{\Phi}_{0} \rangle^{i k} 
\langle\hat{E}\rangle_k^{\ l}  \langle \hat{Y}_u \rangle_l^{\ j}
 + \langle \hat{Y}_u^T \rangle^i_{\ k} \langle\hat{E}^T\rangle^k_{\ l}
 \langle \bar{\Phi}_{0} \rangle^{l j} \right) 
+  {\xi_R} \langle \hat{Y}_\nu^T \rangle^i_{\ k} 
\langle \bar{E} \rangle^{kl} 
\langle \hat{Y}_\nu \rangle_l^{\ j} \right] ,
 \eqno(2.15)
$$
where $\langle \hat{E}\rangle = \langle \bar{E}\rangle= v_E {\bf 1}$.
The form of the first term in Eq.(2.15), $\bar{\Phi}_0 \hat{Y}_u$, 
was first introduced in Ref.\cite{YK_PLB08}. 
The new form (2.15) for  $\langle \bar{Y}_R \rangle$ has been 
adopted in this paper  in order for the  $R$ charge assignment to be
more natural.


Since we deal with mass ratios and mixings only, the common 
coefficients $k_f$, $v_{Sf}$, and so on does not 
affect the numerical results, so that hereafter we omit such
coefficients even if those have dimensions.

\vspace{3mm}

{\large\bf  3 \ $R$ charges and superpotentials}

In this section, we demonstrate how to derive the VEV relations 
presented in Sec.2 from superpotentials given bellow. 
Hereafter, for convenience, we have sometimes 
dropped the notations ``$\langle$" and ``$\rangle$". 

Superpotentials for flavons are determied under 
U(3)$\times$U(3)$'$ symmetries and $R$ charge conservation.  
Once we fix $R$ charge assignment, our superpotentials are 
uniquely  determined without ambiguity.

First, let us show guidelines on the assignment of $R$ charges:

(i) In the Yukawaon model, it is required that all flavons 
take $R$ charges with $R\geq 0$ except for special flavons
$\Theta$ which always take $\langle \Theta \rangle =0$. 
In the conventional models, a U(1) charge Q is usually assumed, 
and thereby, only terms with $Q=0$ are allowed in the Hamiltonian.
When there are terms $A$ and $B$ whose $R$ charges are 
$Q_A=0$ and $Q_B=0$, the combined term $A\cdot B$ is also 
allowed in the Hamiltonian. 
We do not want such the situation. 
Therefore, we require that all flavons take $R\geq 0$, 
and thereby, it is forbidden that a term which already has 
$R=2$ makes an unwelcome higher-dimensional term  
combined with another flavon (or term).
   
(ii) We require that $R$ charges are conserved even under the 
mixing between $f_L$ and $F_L$ (and also between 
$f_R$ and $F_R$) in Eq.(2.4). 
Therefore, flavons given in Eq.(2.4) have to satisfy the
following relations:
$$
R(f_L) =R(F_L) \equiv r_{fL}, \ \ \ \ 
R(f_R) =R(F_R) \equiv r_{fR},
\eqno(3.1)
$$
$$
R(S_f) = R(\Phi_f) =R(\bar{\Phi}_f) = R(\hat{Y}_f) \equiv r_f.
\eqno(3.2)
$$
(Eq.(3.2) does not mean $r_{fL} =r_{fR}$.)
Correspondingly to Eq.(3.1), we require 
$$
R(\bar{A}) = R(A) ,
\eqno(3.3)
$$
for any flavons $A$ and anit-flavons $\bar{A}$. 
Here $R(A)$ denotes $R$ charge of flavon $A$, and so on. 

(iii) Values of $R$ charges should be as possible as 
simple, by taking the basic VEV relations into consideration.
According to the conventional SUSY models, we assign zero 
for the Higgs scalar doublets $H_u$ and $H_d$, e.g. 
$R(H_u) = R(H_d) =0$. 
Our $R$ charge assignment is shown in Table 2.  

Next, let us discuss a simple case which gives 
$b_e = b_\nu =0$ by considering the following superpotential: 
$$
W_S = \sum_{f=e, \nu} \left\{ \lambda_{1f} 
[(S_f)_\alpha^{\ \beta} (P_f)_{\beta k} (\bar{P}_f)^{k\alpha}]
+ \lambda_{2f}  [(S_f)_\alpha^{\ \alpha} ]
[(P_f)_{\beta k} (\bar{P}_f)^{k\beta}] \right\},
\eqno(3.4)
$$
where we have taken $R$ charges of $S_f$ and $P_f$ as
$$
R(S_f) + 2R(P_f) = 2 \ \ \ \ (f=e, \nu).
\eqno(3.5)
$$
The form  (3.4) is newly adopted in this paper. 
(Eq.(3.4) is somewhat improved from the previous paper 
\cite{Yukawaon_seesaw}.)
The vacuum condition for the superpotential (3.4) leads to
$$
S_f = {\bf 1} , \ \ \ P_f \bar{P}_f = {\bf 1}.  \ \ \ 
(f=e, \nu)
\eqno(3.6)
$$
The result $S_f ={\bf 1}$ means $b_f=0$, so that 
$$
\hat{Y}_e \propto \hat{Y}_\nu \propto \Phi_0 \bar{\Phi}_0 . 
\eqno(3.7)
$$

On the other hand, from the relation (2.9), we obtain
$$
R(P_f) = r_f - r_0,
\eqno(3.8)
$$
where $r_0 \equiv R(\Phi_0)$, 
so that, from the $R$ charge relation (3.5), we obtain
$$
r_e = r_\nu = \frac{2}{3} (r_0 +1) .
\eqno(3.9)
$$
However, (3.6) and (3.9) do not mean that $\hat{Y}_e$ and 
$\hat{Y}_\nu$ are an identical flavon.

We take a specific VEV form 
$$
P_f = 
{\rm diag} (e^{i \phi_1^f},  e^{i\phi_2^f}, e^{i\phi_3^f} ),
\eqno(3.10)
$$
from the general form $P_f \bar{P}_f = {\bf 1}$ in (3.6) by assuming 
$\langle  \bar{P}_f \rangle = \langle P_f \rangle^\dagger$ 
and by assuming that the VEV matrix is diagonal.  
Since the VEV matrix form of $S_e$ (and also $S_\nu$) is diagonal,
the VEV matrix $P_e$ is commutable with $S_e$, so that 
the phase parameters $\phi_i^e$ in $P_e$ cannot play any 
physical role in $\bar{P}_e (S_e^{-1}) P_e$.
Therefore, we have simply put $P_f = {\bf 1}$ for $f=e, \nu$
in Eq.(2.14) with Eqs.(2.9) and (2.3). 

For $f=u, d$, we assume the following superpotential:
$$
W_{Pq} = \frac{1}{\Lambda} \left\{ \lambda_{1P} 
{\rm Tr} [P_u \bar{P}_u P_d \bar{P_d} ] + \lambda_{2P} 
{\rm Tr} [P_u \bar{P}_u ] {\rm Tr} [P_d \bar{P_d} ]  \right\} ,
\eqno(3.11)
$$
where we take
$$
R(P_u)+ R(P_d) =1 .
\eqno(3.12)
$$
The SUSY vacuum condition leads to $P_u \bar{P}_u={\bf 1}$
and $P_d \bar{P}_d={\bf 1}$. 

From Eq.(3.12),  we obtain an $R$ charge relation
$$
r_u + r_d = 1+2 r_0 .
\eqno(3.13)
$$
On the other hand, we obtain 
$$
r_u + r_d = r_\nu +r_e = \frac{4}{3} (r_0 +1),
\eqno(3.14)
$$
from the superpotential $W_{\phi}$ given in (4.1).
Eqs.(3.13) and (3.14) fix the value of $r_0$ as
$$
r_0 = \frac{1}{2} , \ \ \ \Rightarrow \ \ \ 
r_e = r_\nu = 1.
\eqno(3.15)
$$

For the VEV relations (2.9) and (2.15),
we consider somewhat tricky prescription: 
We assume existence of $\Theta$ fields which always take 
VEV values $\langle \Theta \rangle =0$.
For example, in order to obtain the VEV relation (2.9),
we assume the following superptential
$$
W_{0} = \mu_{0f} (\Phi_{0f})_i^{\ \alpha}
 (\bar{\Theta}_{0f})_\alpha^{\ i} + 
\lambda_{0f} (\Phi_0)_{ik} (\bar{P}_f)^{k \alpha}
  (\bar{\Theta}_{0f})_\alpha^{\ i} .
  \eqno(3.16)
$$
From $\partial W_0/\partial \bar{\Theta}_{0f} = 0$, we obtain 
VEV relation (2.9).
On the other hand, a derivative of $W_0$ with respective
 to other flavon, for example, $\Phi_0$ leads to 
$\partial W_0/\partial \Phi_0 = \lambda_{0f} \bar{P}_f\Theta_{0f}
=0$. 
However, the result always includes $ \Theta_{0f}$, so that 
the condition does not lead to any new VEV relations. 
This prescription is very useful when one flavon appears 
in the different (two or more) superpotentials.
(For example, instead of $W= (\mu A + \lambda BC) \Theta$, 
we may consider  $W= (\mu A + \lambda BC) (\mu A + \lambda BC)^\dagger$.
However, then we have an $R$ charge constraint $R(A) = R(B)+R(C) = 1$ 
addition to $R(A) = R(B)+R(C)$. Beside, the SUSY vacuum condition
$\partial W/\partial A=0$ will lead to unwelcome relation if 
there is another potential term which includes the flavon $A$.)
Of course, such $\Theta$ flavon prescription is a big ansatz 
in the Yukawaon model. 
We have to search for more reasonable 
prescription in future. 

Similarly, for the VEV relation (2.15), we assume 
the following new superpotential presented in this paper, 
$$
W_R = \left\{ \mu_R (\bar{Y}_R)^{ij} +\frac{\lambda_R}{\Lambda}
\left( (\bar{\Phi}_{0\nu})^i_{\ \alpha} (\bar{P}_d)^{\alpha k}
( \hat{Y}_u)_k^{\ j} + 
(\hat{Y}_u^T)^i_{\ k}  (\bar{P}_d^T)^{k\alpha}
 (\bar{\Phi}_{0\nu})_\alpha^{\ j} 
+ {\xi_R} (\hat{Y}_\nu^T)^i_{\ k} \bar{E}^{kl} 
(\hat{Y}_\nu)_l^{\ j} \right) \right\} (\Theta_R)_{ji} ,
\eqno(3.17)
$$ 
together with
$$
 W_E = \frac{\lambda_{E1}}{\Lambda} [
(E)_{ik} (\bar{E})^{kl} (E)_{lm} (\bar{E})^{mi} ] +
\lambda_{E2} [(E)_{ik} (\bar{E})^{ki}]\, [(E)_{jl} (\bar{E})^{lj} ] ,
\eqno(3.18)
$$
where we have taken 
$$
 R(E)= R(\bar{E}) = \frac{1}{2}.
\eqno(3.19)
$$
Moreover, the form of ${W}_R$, Eq.(3.17) requires a relation
$$
 r_u + r_0 + r_E= 2 r_\nu +r_E , 
\eqno(3.20)
$$
so that we obtain $r_u$ and $r_d$ as follows:
$$
r_u = \frac{3}{2}  , \ \ \ r_d =  \frac{1}{2} .
\eqno(3.21)
$$

Thus, in this section, we have derive the $R$ charges of 
flavons from the desirable superpotential forms. 
In other words, this means that if we start from the $R$ charge
assignment given in Table 2, we can uniquely reach to the desirable 
superpotential forms.    

\vspace{2mm}

{\large\bf  4 \  Relation between $(\phi_1, \phi_2, \phi_3)$ and 
$(m_e, m_\mu, m_\tau)$ }

In this section, we give a relation which connects the phase parameters
$(\phi_1, \phi_2, \phi_3)$ ($\phi_i = \phi_i^u -\phi_i^d$) 
with the family-number-dependent input parameters $(z_1, z_2, z_3)$. 
Although the basic idea has already given in Ref.\cite{K-N_PRD15}, 
the explicit relation is renewed in the present paper as follows: We consider 
the following superpotential in this paper.
$$
W_{\phi} = \left\{ \lambda_1 \left[
(P_u)_{i\alpha} (\bar{P}_d)^{\alpha j} +
(P_d)_{i\alpha} (\bar{P}_u)^{\alpha j} \right] 
+\lambda_2  \left[
(P_\nu)_{i\alpha} (\bar{P}_e)^{\alpha j} +
(P_e)_{i\alpha} (\bar{P}_\nu)^{\alpha j} \right] 
+ \lambda_3 (\Phi_0)_{ik} (\bar{\Phi}_0)^{kj}
\right\} (\hat{\Theta}_{\phi})_j^{\ i} .
\eqno(4.1)
$$
In order to get Eq.(4.1), it is essential that the VEVs of the 
flavons satisfy the following $R$ charge relation 
$$
R(P_u) +R(P_d) = R(P_\nu) +R(P_e) = 2 R(\Phi_0) .
\eqno(4.2)
$$

The first term in Eq.(4.1) gives a VEV relation 
$$
\left[
(P_u)_{i\alpha} (\bar{P}_d)^{\alpha j} +
(P_d)_{i\alpha} (\bar{P}_u)^{\alpha j} \right]
\propto \cos\phi_i ,
\eqno(4.3)
$$
where $\phi_i = \phi_i^u -\phi_i^d$.
On the other hand, VEVs of the second and third terms are
proportional to $1$ and $z_i^2$, respectively. 
Therefore, we obtain a relation
$$
\cos\phi_i = a + b z_i^2 ,
\eqno(4.4)
$$
where the parameters $a$ and $b$ are family-number-independent
parameters.

Note that observable parameters in the three phase parameters  
$(\phi_1, \phi_2, \phi_3)$ are only two. 
When we denote 
$$
\phi_1 = \phi_0 + \tilde{\phi}_1, \ \ \ 
\phi_2 = \phi_0 + \tilde{\phi}_2, \ \ \ 
\phi_3 = \phi_0 ,
\eqno(4.5)
$$
the parameter $\phi_0$ is not observable. 
Therefore, we can always choose arbitrary value of $\phi_0$, 
so that  the relation (4.4) is satisfied by choosing 
two family-number-independent parameters $a$ and $b$ 
suitably. 
(Note that although the parameter $\phi_0$ is not observable 
in the framework of the standard model (SM), the parameter 
$\Phi_0$ in the Yukawaon is observable because we consider
U(3)$\times$U(3)$'$ which are gauged. 
The value $\phi_0$ will be confirmed by future experiments.)

Explicitly, we can obtain numerical results as follows:
By eliminating the parameter $a$, we obtain a relation 
$$\cos\phi_1 -b\, z_1^2 = \cos \phi_2 -b\, z_2^2 
= \cos \phi_3 -b\, z_3^2.
\eqno(4.6)
$$
Then, we can obtain a relation for the parameter $\phi_0$: 
$$
b=\frac{\cos \phi_3 -\cos \phi_1}{z_3^2 -z_1^2} =
\frac{\cos \phi_3 -\cos \phi_2}{z_3^2 -z_2^2} .
\eqno(4.7)
$$
Since we have obtained the parameter values \cite{Yukawaon_seesaw}
$$
(\tilde{\phi}_1, \tilde{\phi}_2) = (-176.05^\circ, -169.91^\circ) ,
\eqno(4.8)
$$
from fitting of the observed CKM mixing data,
we obtain family-number-independent parameters $(a, b)$ 
$$
 (a,b) = (1.71573, -0.790018) ,
\eqno(4.9)
$$
together with a value $\phi_0 = 33.905^\circ$. 
Note that the value $\phi_0 = 33.905^\circ$ is observable if we 
consider U(3) family gauge bosons, although it is not observable
in the CKM parameter fitting. 
We predict the phase parameters $(\phi_1, \phi_2, \phi_3)$ as 
follows:
$$
(\phi_1, \phi_2, \phi_3)=(-142.14^\circ, -136.00^\circ, 33.91^\circ) .
\eqno(4.10)
$$
In future, those values will be confirmed by family gauge boson experiments.

\vspace{2mm}

{\large\bf  5 \   Concluding remarks}


In conclusion, we have given formulation of a new 
Yukawaon model on the basis of seesaw type mass matrix model
by presenting U(3)$\times$U(3)$'$ assignments, $R$ charges of 
flavons, superpotential forms, and so on. 
In spite of a model with quite few parameters, the model can give a 
remarkable agreement with the observed quark and lepton mixings 
and mass ratios.
Those phenomenological (numerical) results of the model have already 
been reported in the previous paper \cite{Yukawaon_seesaw}. 
We emphasize that the phenomenological success
highly depends on whether we can assign the $R$ charges
reasonably and consistently or not.  
It is in the present paper that the explicit 
$R$ charge assignment is completed.

The phenomenological success of the present model  seems to suggest the following 
points:

\noindent (a) The observed quark and lepton masses and mixings 
are caused by a common origin.

\noindent (b)  Flavor physics should be investigated on a flavor 
basis in which charged lepton mass matrix is diagonal.
(Mass matrix of family gauge bosons is also diagonal 
in this basis, so that, family gauge bosons will not cause flavor violation 
in the charged lepton sector \cite{Sumino_PLB09} .)

\noindent (c) Masses and mixings in the quark sector are 
given by the parameters $b_u$ and $b_d$ in the form of $S_f$ 
as shown in Eq.(2.7).
This mechanism is very interesting. 

On the other hand, for the theoretical aspect, the model 
has still many problems which should be improved in future.
For example, in the present paper, we did not discuss explicit scales of
$\Lambda$ and $\Lambda'$, 
although we have tacitly assumed that $\langle A_{ij}\rangle
\sim \Lambda$, $\langle A_{\alpha\beta}\rangle \sim \Lambda'$
and $\langle A_{i\alpha}\rangle \sim \sqrt{\Lambda\Lambda'}$.     
The choice is highly correlated in the tininess of neutrino 
masses. 
Since we have discussed masses and mixings only, 
we have neglected the common coefficients and VEV values 
in the sectors, for example, $k_f$ in Eq.(2.3)
$v_{Sf}$ in Eq.(2.7), $v_0$ in Eq.(2.10), $k_R$ in Eq.(2.15), and so on. 
 Those are our future task. 

We believe that the present model can give fruitful 
suggestions for the study of flavor physics.


\vspace{2mm}

{\large\bf  Acknowledgement}

This work is supported by JSPS (Grant No. 16K05325).

\newpage
\vspace{5mm}
%

\end{document}